\documentclass{bmcart}
\usepackage{url}
\usepackage{amsmath,amssymb}
\usepackage{graphicx}
\usepackage{subfig}

\begin{document}
	
\begin{frontmatter}
		
\begin{fmbox}
\dochead{Research}
			
\title{The collective vs individual nature of mountaineering: a network and simplicial approach}
\author[
addressref={aff1},                   
email={s.krishnagopal@ucl.ac.uk}   
]{\inits{S.K.}\fnm{Sanjukta} \snm{Krishnagopal}}

\address[id=aff1]{
	\orgdiv{Gatsby Computational Neuroscience Unit},             
	\orgname{University College London},          
	\city{London-W1T 4JG},                              
	\cny{UK}                                    
}

\end{fmbox} 

\begin{abstractbox}
	
\begin{abstract}
 Mountaineering is a sport of contrary forces: teamwork plays a large role in mental fortitude and skills, but the actual act of climbing, and indeed survival, is largely individualistic.
 This work studies the effects of the structure and topology of relationships within climbers on the level of cooperation and success. It does so using simplicial complexes, where relationships between climbers are captured through simplexes that correspond to joint previous expeditions with dimension given by the number of climbers minus one and weight given by the number of occurrences of the simplex.
 First, this analysis establishes the importance of relationships and shows that chances of failure to summit reduce drastically when climbing with repeat partners. From a climber-centric perspective, it finds that climbers that belong to simplexes with large dimension were more likely to be successful, across all experience levels.
 Then, the distribution of relationships within a group is explored to identify collective human behavior determining expedition styles: from polarized to cooperative. Expeditions containing simplices with large dimension, and usually low weight, implying that a large number of people participated in a small number of joint expeditions, tended to be more cooperative, benefiting all members of the group, not just those that were part of the simplex. On the other hand, the existence of small, usually strong, subgroups lead to a polarized style where climbers that were not a part of the subgroup were less likely to succeed.
 Lastly, this work examines the effects of individual features (such as age, gender, experience etc.) and expedition-wide factors (number of camps, total number of days etc.) that may play bigger roles in individualistic and cooperative expeditions respectively. Centrality indicates that individual traits of youth and oxygen use while ascending are strong drivers of success. Of expedition-wide factors, the expedition size and number of expedition days are found to be strongly correlated with success rate.

\begin{keyword}
	\kwd{mountaineering data analysis}
	\kwd{social network analysis}
	\kwd{simplicial complex}
	\kwd{group dynamics}
	\kwd{success on Everest}
\end{keyword}

\end{abstract}

\end{abstractbox}
%

\end{frontmatter}

\section*{Introduction}

Extreme mountaineering is an increasingly popular sport that straddles the boundary between an individual sport and a group activity. Under extreme settings, success requires not only very high levels of physical fitness \cite{huey2001,szymczak2021}, but also psychological \cite{ewert1985} and sociological state \cite{savage2020personality,helms1984}. A majority of mountaineering is undertaken as a part of an expedition, with a high level of inter-dependency between climbers, often relying on their fellow climbers in life or death scenarios. The psychology is largely driven by relationships between climbers; for instance climbers that frequently climb together may develop better communication and group dynamics, and consequently lower failure. Simultaneously, certain aspects of extreme mountaineering are well-known to be individualistic, especially as one gets closer to the death zone (8000m altitude) \cite{crockett2020self}. The interplay of these conflicting forces makes extreme mountaineering an interesting setting to study collective human behavior. Such analyses are made possible using data from the meticulously documented Himalayan dataset \cite{data04}.This work is a data-driven study of the structure of relationships between climbers within an expedition and its effects on individual and group success.


Relationships between climbers in an expedition may result from a previous joint expedition comprising of a subset of the climbers. Relationships between climbers are naturally captured by a network framework. A conventional network can capture the interactions between two climbers (or `nodes') as a link (or `edge') with a weight given by the number of expeditions that they jointly participated in. Network approaches have been used successfully in predictive medicine \cite{krishnagopal2021}, climate prediction \cite{steinhaeuser2011}, predictions in group sports \cite{lusher2010}, disease spreading \cite{disease2019} etc. However, such pairwise networks are unable to accurately capture scenarios where interactions between more than two climbers occur, and reduce them to multiple pairwise interactions, which is fundamentally misleading. Simplicial complexes \cite{bassett,bianconi2021higher,battiston,tina} are an important tool for modeling systems with simultaneous interactions between more than two entities.
An expedition involving a subgroup of climbers climbers can, for instance, be represented as a filled triangle, differentiating it from a set of three edges. The weight of this filled triangle, which is a simplex, is given by the number of previous expeditions containing the three climbers.
Indeed, simplicial representations involving simplexes (e.g. triangles, tetrahedra etc.) have been successfully used to model a variety of systems such a social communication networks \cite{wang2020social}, complex systems \cite{SAL18,BEN16}, disease spreading \cite{IAC19} etc. and are a rapidly growing field of data analysis. 


The composition of a group and the relationships therein influence its effectiveness. \cite{sherman2013national} uses the distribution of climbers' nationalities as a meaningful measure of extent of collaboration and competition, and finds that collective mentality, which may often result from strong relationships, boosted summiting when national diversity was high. Despite its wide ranging applications, there is limited investigation of the effects of relationships on collective behavior. This work studies how the diversity and structure of relationships, formed through climbing together, influence the cooperation or competition between them and ultimately success.
Specifically, it uses topological relationships between climbers as a predictor of expedition style on a spectrum from polarized (where subgroups with strong relationships and without relationships (or weak relationships) are likely to have different outcomes) to globally cooperative (where climbers, regardless of their relationships with others, share the benefit of the existence of high-dimensional simplexes in the group). 

Various other factors, both personal and expedition-wide, have differing effects on success based on expedition style, with the former playing a larger role in individualistic expeditions and the latter in a cooperative. Personal elements such as effective use of proper equipment, climber experience, mental state etc. \cite{schussman1990}  are crucial to maximizing safety and chances of success. Several works have studied the effects of age, sex, nationality \cite{huey2007,weinbruch2013},  experience \cite{huey2020}, commercialization \cite{westhoff2012} etc. on success, and highlight the importance of age as a dominant determining factor. Expedition-related factors such as effective use of proper equipment, climber experience, mental strength and self-reliance are all measures \cite{schussman1990} to increase safety and chances of success. In, \cite{krishnagopal2021success}, the effects of various personal features like age, sex, experience etc. as well as expedition-wide factors like length of expedition, number of sherpas etc. on success are studied using a multiscale network. 
This work investigates relationships between climbers and personal features such as oxygen use, age, sex, experience etc. through a bipartite network, projected into factor-space for further analysis \cite{krishnagopal2020,larremore2014}. The natural question emerges, which of these features, which can be represented by nodes, are central to maximizing chances of success? An active area of research investigates the importance of nodes \cite{mo2019} through centrality-based measures \cite{sola2013,saito2016} that serve as reliable indicators of `important' features. 
Expedition-wide factors such as ratio of sherpas to paying climbers, number of days to summit, and number of camps, intra-expedition social relationships etc. are also considered as predictors of success. Outside of \cite{krishnagopal2021success}, which this works builds on, there is limited literature that studies the effect of such expedition-wide factors. 

\section{Data}

The data used comes from the open access Himalayan Database \cite{data04}, which is a compilation of records for all expeditions from 1905 to 2021 in the Nepal Himalayan ranges. The dataset has records of 468 peaks, over 10,500 expedition records and over 78,400 records of climbers, where a record of any type is associated with an ID. 
Details of what 
The following data are extracted from the expedition records:
\begin{itemize}
	\item Peak climbed (height).
	\item Days from basecamp to summit.
	\item Number of camps above basecamp.
	\item Total number of paying members and hired personnel.
	\item List of all joint expeditions that contain a set of climbers (calculated)
	\item Result: (1) Success (main peak/foresummit/claimed), (2) No summit, (3) Death
\end{itemize}
The success rate of an expedition is calculated as the fraction of members that summited. 
The following data are extracted from climber records:

\begin{itemize}
	\item Demographics: Age, Sex, Nationality.
	\item Oxygen use: ascending or descending.
	\item Previous experience above 8000m (calculated).
	\item Result: 
	\begin{enumerate}
		\item Success
		\item Altitude related failure: Acute Mountain Sickness (AMS) symptoms, breathing problems, frostbite, snowblindness or coldness.
		\item Logistical or Planning failure: Lack of supplies, support or equipment problems, O2 system failure, too late in day or too slow, insufficient time left for expedition.
		\item Fatigue related failure: exhaustion, fatigue, weakness or lack of motivation.
		\item Accident related failure: death or injury to self or others.
	\end{enumerate}
The climber ID of each climber is tracked in the dataset through history to generate a log of their previous expeditions. The number, type, and nature of previous joint expeditions are then extracted by calculating the overlap between climber logs. 
\end{itemize}

\section{The effect of climbing with repeat partners}

\begin{figure}[htbp!]
	\centering
	\includegraphics[width=0.7\linewidth]{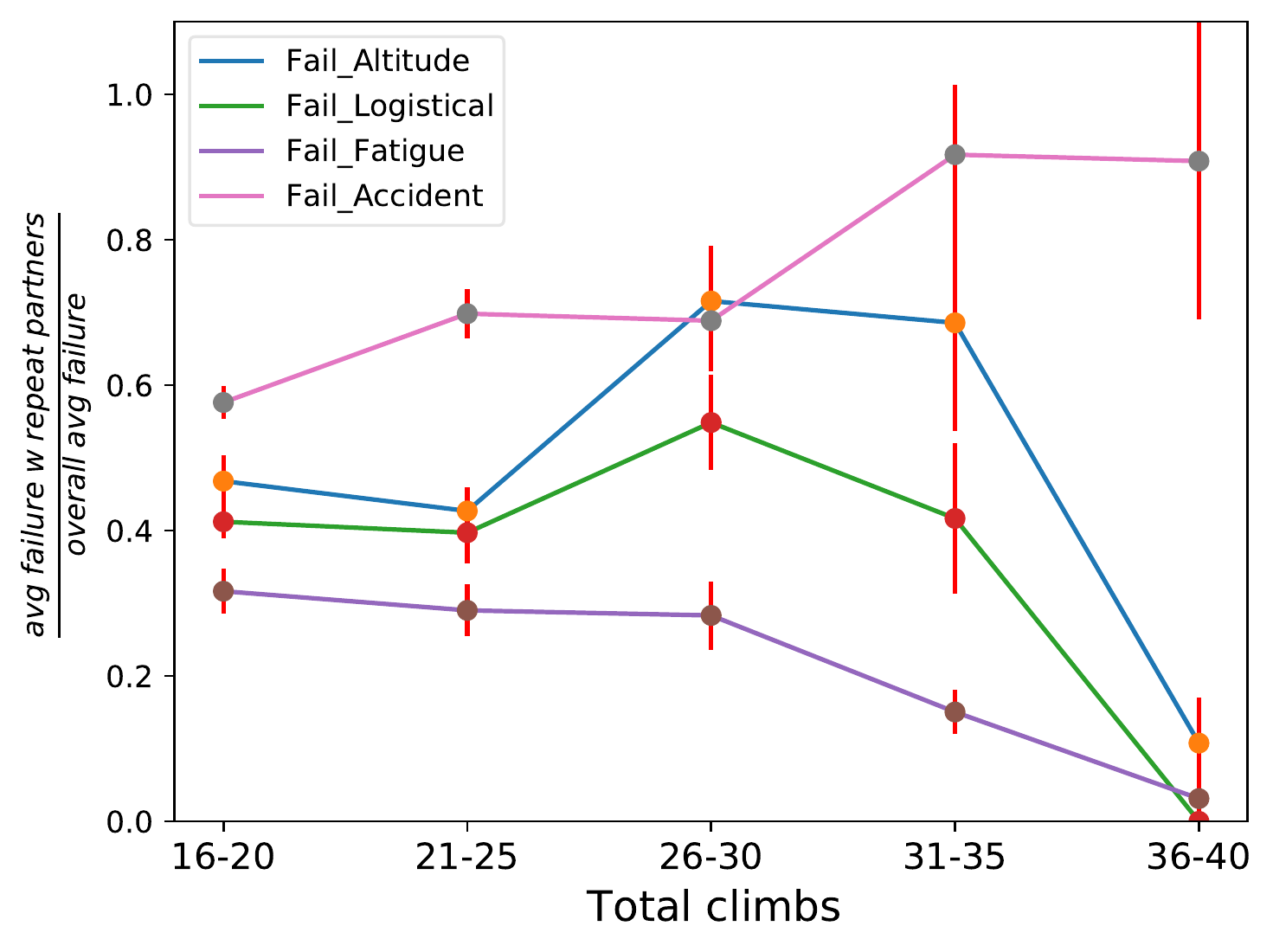}
		\caption{The fraction of failures (of several types) when climbing in a group with at least one repeat partner (someone they have done a logged Himalaya expedition with before) over the their personal average. The x-axis denotes the total number of previous climbs to normalize for levels of experience.}
\label{fig:friends}
\end{figure}

The first question is, is it advantageous to climb with people one has climbed with before? It is natural to assume that friendship and familiarity with each others' climbing style may improve confidence and the accuracy of calculated risks, thus limiting failure, but may also lead to a false and potentially dangerous sense of comfort. Here, the average failure rate of a climber is compared with their failure rate when climbing in expeditions with at least one repeat partner (that they have climbed with before). Fig.\ref{fig:friends} shows the fraction of failures when climbing with repeat partners over the climber average. These failures are classified into altitude related, fatigue related, logistical and planning failures, and accident/illness. The effect of total experience is normalized for by plotting across the total number of climbs on the x-axis starting at least 15 climbs, hence not considering beginner climbers.

The y-axis in Fig.\ref{fig:friends} plots the ratio of failure rate when climbing with at least one repeat partner over average failure rate over all climbs. Since the values of all failure ratios are below 1, the chance of failure is significantly lower when climbing with repeat partners, in every failure category. In particular, the probability of failure due to fatigue-related issues is the most decreased when climbing with repeat partners, followed by failure due to logistical or planning issues. This may be expected since climbers that have participated in previous joint expeditions typically are better at communication and knowing each others' limitations. Note that only climbers with over 15 logged climbs are considered, indicating that complete lack of experience is not a cause of failure. Additionally, the most experienced climbers (that have logged 36-40 climbs) have nearly no failure due to fatigue or logistics, as one may expect. Similarly, failure due to altitude-related and cold-related problems also drastically reduce when climbing with repeat partners. Lastly, the cause of failure due to accident shows an increasing trend as a function of increasing experience, which may be attributed to the fact that more experienced climbers tend to tackle more dangerous mountains.

\section{Methods}
\subsection{Simplicial complexes}
This work is interested in analyzing the \textit{structure of relationships} between subgroups of climbers and their effect on expedition style; from an individualistic to cooperative spectrum. Relationships between climbers may result from previous joint expeditions.
Such joint expeditions may involve a subgroup of individuals of size two or more. 
Networks, although natural to model interactions, can only capture pairwise relationships and hence fail to accurately capture higher-order interactions (interaction between more than 2 climbers in a single expedition). However, multi-node interactions can be explained by a higher-order network: a mathematical framework called {\em simplicial complexes}. For instance, a three-way interaction can be represented as a triangle, four-way as a tetrahedra etc.

\begin{figure}[htbp!]
\includegraphics[width=0.7\columnwidth]{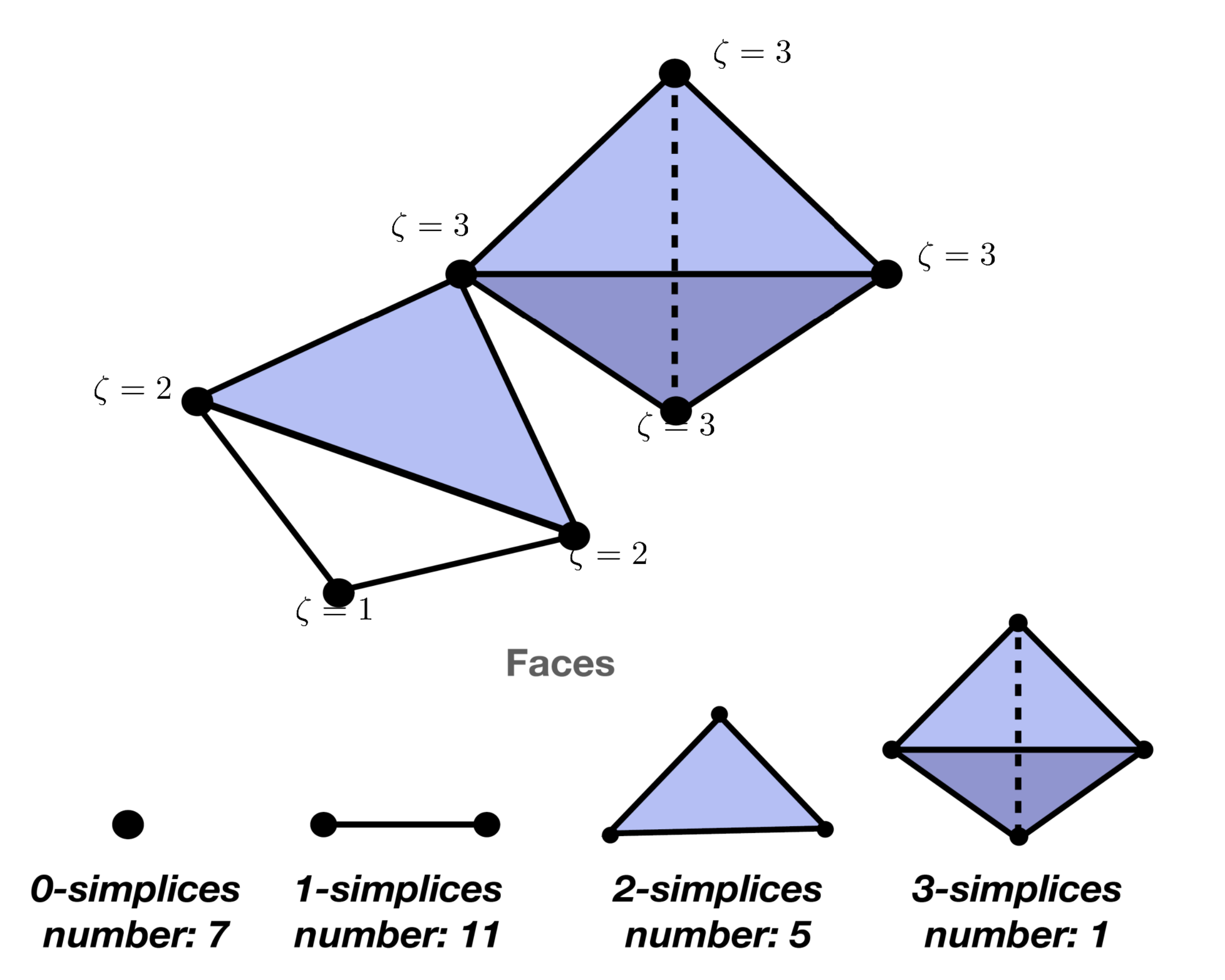}
	\caption{simplexes that form the faces of the simplicial complex. The number of $k$-simplexes in the top simplicial complex are listed. The influence, or maximal simplicial dimension, $\eta$ for each node is listed.}
\label{faces}
\end{figure}

Mathematically, given a set of $l$ nodes ${n_0, n_1 \ldots, n_l} \in N$ in a network, a $p-$simplex is a subset $\sigma_p = [{n_0,n_1,\ldots, n_p}]$ of $p$ nodes and a $q-$face of  $\sigma_p$ is a set of $q$ nodes (for $q<p$) that is a subset of the nodes of  $\sigma_p $.
A \emph{simplicial complex} ${K}$ consists of a set of simplexes, that are closed under inclusion:
\begin{equation}
\tau\subseteq\sigma \Rightarrow \tau \in {K} \text{~for any}\ \sigma\in {K},
\end{equation}
where `$\subseteq$' denotes the subset relation between $\sigma$ and $\tau$, two subsets of the simplicial complex. When $\tau\subseteq \sigma$, we say that $\tau$ is a \emph{face} of $\sigma$, which by the inclusion axiom implies \emph{every face of a simplex is again a simplex}. Fig.~\ref{faces} shows examples of faces of a simplicial complex. 

The \emph{dimension} of a simplex equals the number of vertices in the simplex minus one; for instance $0$-dimensional simplexes are nodes and $1$-dimensional simplexes are edges. Each previous joint expedition is represented as a simplex. The dimension of a simplicial complex is the largest dimension of its simplexes, where each expedition is modeled as a simplicial complex with the nodes representing climbers.
By ${S}_k$ we will denote the  set of $k$-dimensional simplexes, i.e. as
\begin{equation}
{S}_k := \lbrace \sigma\in S : \vert \sigma\vert = k+1\rbrace,
\end{equation}
We call the simplexes in ${S}_k$ the $k$-simplexes of ${K}$ and let $|S_k|$ denote the number of $k$-simplexes in the simplicial complex. 

If three climbers $i,j,k$ in an expedition have participated in an expedition previously, this is represented as a 2-simplex (triangle) with nodes $i,j,k$ as its $0-$faces. Each climber can be a part of multiple previous expeditions, and hence be the face of multiple simplexes. Let the simplexes (denoted by $\sigma$) that contain individual $i$ be given by: $n_i \in \sigma_{k1}, \sigma_{l2}, \ldots, \sigma_{k}$ where $\sigma_k$ is a $k-$simplex such that their simplicial dimensions are ordered: $k1 \leq k2 \leq \ldots \leq k$. In other words, we order the simplexes that a climber belongs to in order of their simplicial dimension. 

The influence of a climber is related to number of other climbers they are connected with and whether those relationships coevolved. The \textit{influence} $\zeta_i$ of the $i^{th}$ climber is defined to be the maximum number of people they have climbed with in a single expedition, i.e., dimension of its largest simplex $\zeta_i = k$. Each climber is associated with an influence value. For a climber who has no previous expeditions with any other climbers $\zeta_i=0$. The co-influence of an expedition $E$ is given by the the average influence of all climbers.

\subsection{Topological data analysis}

Topological Data Science(TDA) is a new and rapidly growing subfield in machine learning and data science for the analysis of simplicial complexes \cite{wasserman2018topological}. The central assumption of TDA is that complex and high dimensional data has an underlying shape captured by topological descriptors, which can be exploited for its analysis. Commonly used topological descriptors are simplexes as well as Betti numbers, where the $k^{th}$ Betti number gives the number of $k-$dimensional cavities in the simplicial complex. This work derives inspiration from persistence homology \cite{aktas2019persistence}, the primary data analysis methodology in TDA, which attempts to extract topological descriptors (e.g. simplexes) in the data that persist over various threshold values. 
Conventionally, topological features are recorded by creating persistence diagrams. These diagrams plot the birth time (on the x-axis), and death time (on the y-axis) of a simplex where time is measured through changing a threshold parameter. The expectation is that topological features that persist for a large amount of time or across various thresholds are features that persist and hence are integral to the analysis of the simplicial complex.

In a similar vein, this work assigns a weight to each simplex in an expedition, given by the number of previous expeditions of the subgroup represented by the simplex, i.e., the number of times the simplex occurs $w_{\sigma} = \# \sigma$. For example, if climbers $i,j,k$ have had four previous joint expeditions together, then the simplex $\sigma_2 = [{n_i,n_j,n_k}]$ is a $2-$simplex with a weight $w_{\sigma_2} =4$. Naturally the higher the weight of the simplex, the stronger the multi-node relationships between the individuals. One can threshold the weight during generation of the simplicial complex such that only simplexes with weights larger than the threshold remain, i.e., only relationships stronger than the threshold are captured. Varying the threshold over a range of values, persistent simplexes (that persist across various weight thresholds $\tau$). One can then study properties of the simplicial complex across $\tau$, such as evolution of the number, distribution and dimensionality of simplexes, as well as the nature of persistent simplexes on outcomes.

\section{Influence as a predictor of climber success: correlating simplicial dimension}
\label{sec:simp_dim}

\begin{figure}[htbp!]
	\includegraphics[width=0.8\columnwidth]{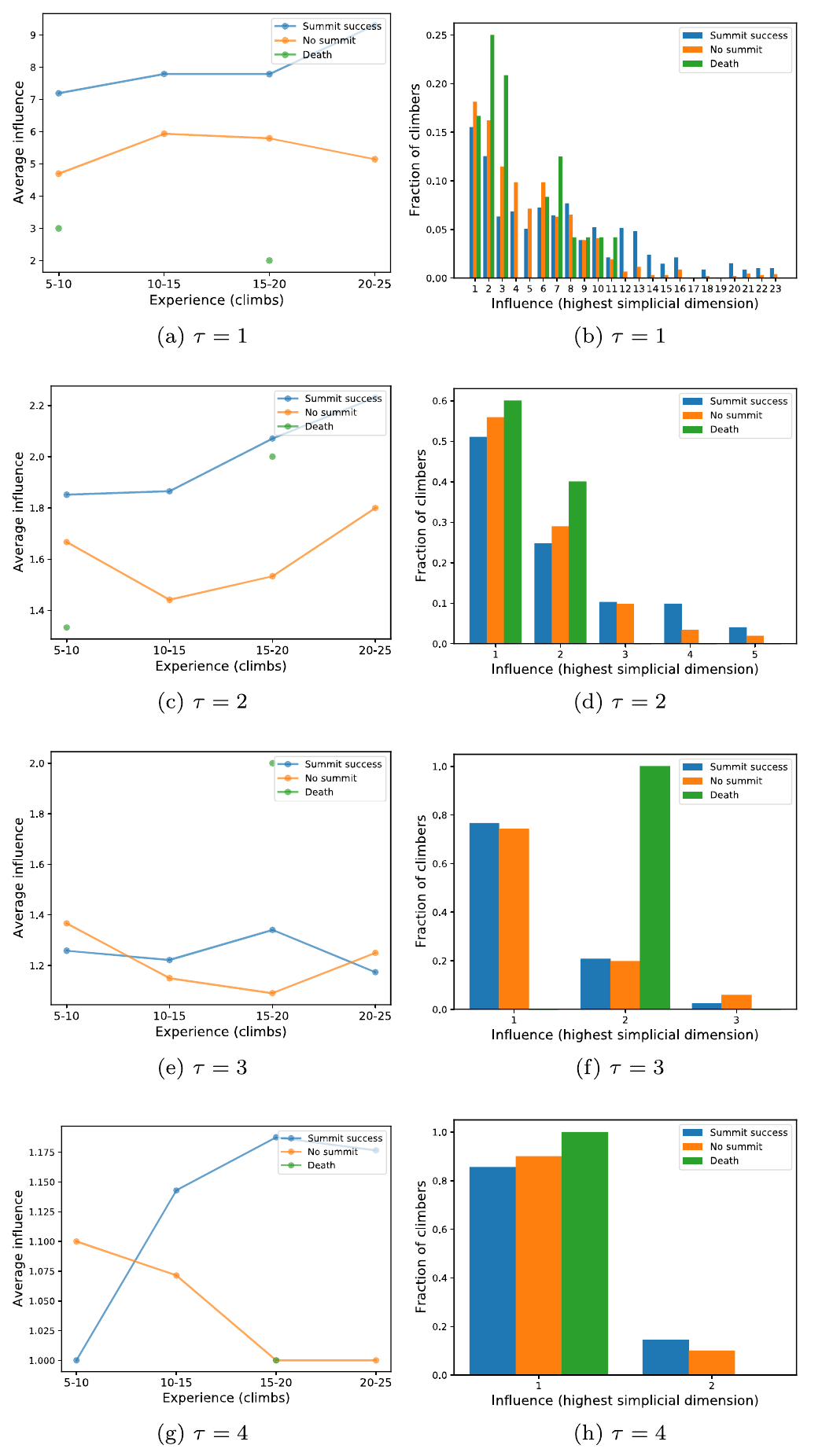}
	\caption{(a,c,e,g) Comparison of average influence (dimension of largest simplex) of climbers falling under three categories: summit success, failure, and death. Statistics are normalized by total previous experience (logged climbs) on the x-axis. (b,d,f,h) Corresponding histogram of the distribution of influence (highest simplicial dimension) for the three categories normalized by the total number of individuals in each category. Total number of deaths are in single digits across all panels. Climbers with zero influence are excluded to yield a total of $n=3051$ climbers for $\tau=1$, $n=731$ for $\tau=2$, $n=342$ for $\tau=3$, and $n=182$ for $\tau=4$. The panels on different rows are across different simplicial weight thresholds $\tau$.}
	\label{fig:climber_across_thresholds}
\end{figure}

Climbing with people that one has relationships with naturally plays a role on the expedition. Here we investigate the precise nature of that role. Fig. \ref{fig:climber_across_thresholds} (left) shows the average influence or simplicial dimension of the largest simplex averaged across climbers from three categories: summit success, no summit, and death. The panels from top to bottom corresponding to increasing the simplicial threshold $\tau$ where only simplexes with a weight equal to or greater than the threshold (stronger relationships) persisted. As seen across all panels, successful climbers almost always had a significantly higher influence than failed climbers, independent of experience. As simplicial threshold increases, the number of simplexes that are above the threshold decrease. Simplexes with large dimension are particularly infrequent and hence less likely to persist. This is observable through the decrease in average influence $\eta$ over increasing $\tau$. The number of individuals that belong to a simplex of dimension 1 or larger also decrease, and hence variances are larger. The number of deaths were in single digits and hence statistically insignificant. Fig. \ref{fig:climber_across_thresholds} (right) shows the distribution of the highest simplicial dimension for increasing threshold $\tau$ from top to bottom. The histogram values are normalized by the total number of individuals in that category for comparison across categories such that the sum of all values in a given category is one. As seen in all panels, successful climbers are relatively more likely to have higher of influence, whereas no-summit cases were relatively more likely to have lower values of influence. Again, climber size reduces with increasing $\tau$. The number of individuals considered are a total of $n=3051$ climbers for (a,b) $\tau=1$, $n=731$ for (c,d) $\tau=2$, $n=342$ for (e,f) $\tau=3$, and $n=182$ for (g,h) $\tau=4$.
Thus find that the highest simplicial dimension is correlated with climber success, i.e., with large values of the highest simplicial dimension are more likely to be successful.


\section{Classification of expedition style through persistent simplexes}

Section \ref{sec:simp_dim} found that climbers with higher influence tended to be more successful. This section investigates how the distribution of influence affects expedition strategies. How can the topology and strength of relationships within an expedition be used to predict the strategy of the group, i.e., the spectrum between polarized vs globally cooperative? Polarized expeditions have sub-groups within the expedition that tend to perform differently, indicating a more individualistic strategy. Globally cooperative expeditions tend to display a more cooperative strategy with uniform outcomes largely independent of climber influence.

\begin{figure}[htbp!]
	\includegraphics[width=0.95\textwidth]{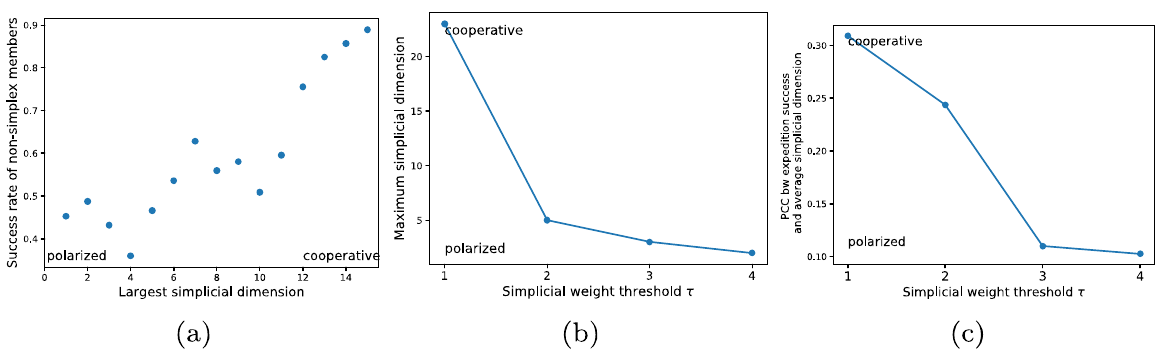}
	\caption{(a) Average success rate of members of expeditions that were not part of the largest simplex as a function of the dimension of the largest simplex. Error bars are plotted in red.  (b) Maximum simplicial dimension of all simplices in the expedition as a function of simplicial weight threshold $\tau$. (c) Pearson's correlation coefficient between expedition success rate calculated over all members and average simplicial dimension of all simplexes in the expedition as a function of $\tau$. Expeditions with no simplex of dimension greater than zero are excluded to yield a total number of expeditions of $n=273$ for $\tau=1$, $n=273$ for $\tau=2$, $n=173$ for $\tau=3$, and $n=118$ for $\tau=4$.}
	\label{fig:exp_style}
\end{figure}

Fig. \ref{fig:exp_style}(a) shows the success rate of members of the expedition that are not part of the largest simplex as a function of the simplicial dimension of the largest simplex in the expedition. Success rate of the `outside' members is defined as the fraction of individuals that did not belong to the largest simplex that reached the summit and back. This provides insight on the expedition style and its effect on climbers that did not belong to the largest subgroup. 

As seen in the figure, outsider success rate is positively correlated with maximum simplicial dimension, indicating that expeditions with large maximal dimension tended to be more cooperative, benefiting not only climbers that were faces of the largest simplex, but also those that did not belong to the largest simplex. Whereas, expeditions where the largest simplex had relatively small dimension resulted in low success rates between non-simplex members, despite having high likelihood of success, irrespective of simplicial dimension, of simplex-members as seen in \ref{fig:climber_across_thresholds}. 
Fig. \ref{fig:exp_style}(b) shows that the maximum simplicial dimension in the expedition is inversely correlated with the simplicial weight threshold $\tau$. Simplexes with large dimension were less persistent. In other words, subgroups within an expedition comprising of a large number of individuals tended to have weaker relationships (existed for low $\tau$) and be more cooperative, benefiting all members of the group. On the other hand, the existence of small but strong subgroups (that persisted for large $\tau$) that had previous climbed together several times lead to a polarity in the group dynamics, leading to a more polarized style where climbers that were not a part of the subgroup were not benefited and tended to be less successful. Statistics are averaged over all levels of experience. Lastly, Fig. \ref{fig:exp_style}(c) validates our results with the insight that the correlation between the total expedition success rate (calculated across all members) and average simplicial dimension is stronger for low $\tau$, i.e., in the cooperative regime, whereas in the polarized regime average influence becomes a poor predictor of success.

Our analysis reveals that strong previous relationships within an expedition are found when the size of the subgroup is small and largely benefit only members of the subgroup leading to subgroups with different likelihoods of success. However the existence of weak relationships between a larger fraction of the expedition tends to unite the team and result in higher homogeneity in their success.

\section{Polarized vs cooperative expeditions: other factors determining success}
\label{sec:centrality}

The relationships within an expedition serve as a predictor of the extent of cooperation and competition. However several other factors, both personal and expeditional, play a role in success to different degrees based on the expedition style.
For instance, climber-specific factors may play a larger role in expeditions that fall on the individualistic side of the style spectrum, whereas expedition-wide factors that are shared across the expedition may play a larger role in expeditions that fall on the cooperative side of the style spectrum.

\subsection{Individual features}
\label{sec:personal}
Various aspects of mountaineering, such as physical fitness and skill, are indeed unconditionally personal.
Here, we study the importance of the following  $d = 6$ personal features: age, sex, nationality, experience above 8000m, oxygen use while ascending and oxygen use while descending. To avoid biases originating from differences in the mountain, data from only one mountain (Mount Everest) is considered. Expeditions with less than 12 climbers are excluded, as are expeditions that resulted in death. An individual is only considered successful is they satisfy both the following criteria: summit and safe descent without requiring rescue. 
To generate the intra-expedition network, we start with a bipartite network $P$ between climbers and features, where a climber is connected (using binary weights) to the features that they are affiliated with. A climber is connected to the `sex' node if they are male, and age is binarized into above and below median age $(40)$. 

\begin{figure}[htbp!]
	\includegraphics[width=0.95\columnwidth]{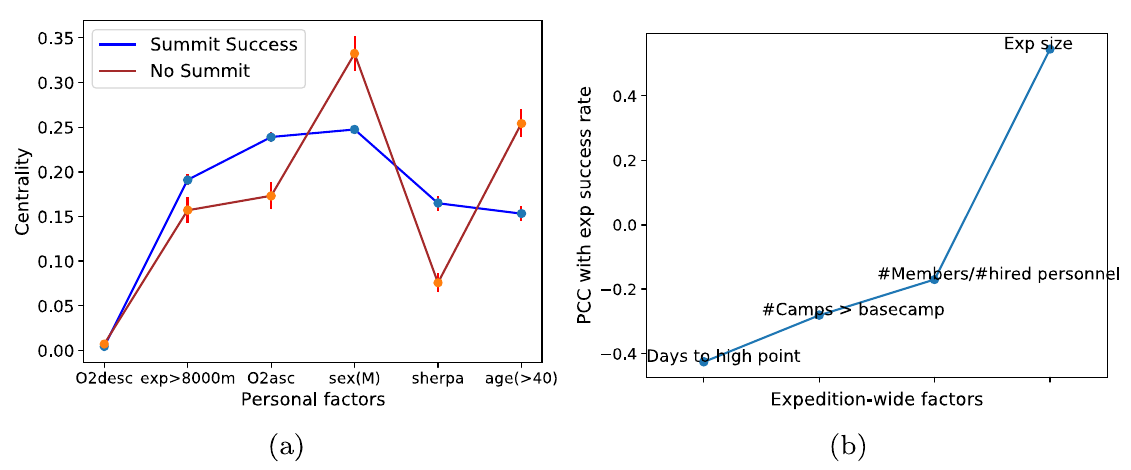}
\caption{(a) Mean eigenvector centrality as a function of expedition features for Everest expeditions greater than 12 members plotted for groups of successful vs unsuccessful climbers ordered by increasing difference between success and no-success centralities. Error bars show standard error on the centrality. (b) Pearson's correlation coefficient (PCC) between layer (factor) values and expedition success rate. The exact values across x-axis layers are $-0.45, -0.36, -0.12, 0.57, 0.84$. The corresponding $p$-values are $5.5 \times 10^{-10}, 1.15 \times 10^{-6}, 0.1, 5.7 \times 10^{-16}, 8.9 \times 10^{-47}$.}
\label{fig:centrality}
\end{figure}

We then generate an intra-expedition network $I$ of size $d \times d$ by projecting the bipartite network into feature space as follows: $I = P^T P$. The edge weight between two nodes (features) is given by the number of people that are connected to both the features. Since the feature graph $I$ is a direct projection of the bipartite graph, it encodes similarities between two features calculated through their simultaneous co-expression in the climbers. We do this independently for all successful and unsuccessful individuals, generating two different networks that encode the properties of the successful and unsuccessful sets of climbers in their network structure.

To explore such properties, measures such as centrality provide insight into the importance of different features \cite{mo2019}. For instance, if the group were comprised of mostly high-age individuals, the centrality of the `age' node would be relatively high. The eigenvector centrality \cite{sola2013} is a commonly used measure of how central each feature is in a given graph. Naturally, \textit{differences} in feature centrality between successful vs unsuccessful sets of climbers is a measure of how important the feature may be for success. 

As seen in Fig. \ref{fig:centrality} (a), where the x-axis is ordered in increasing order of difference in centrality, the least central feature in determining success on summit was the use of oxygen while descending, which is expected since descent features have no effect on summit prospects, except for indicating that oxygen was available on descent meaning there wasn't excessive use during ascent. It is worth noting that most fatalities on Everest happen during the descent.
The next features that were slightly more central in successful summits were previous experience above 8000m (for reference, Everest is at 8849m), followed by use of oxygen while ascending. Importantly, there are few studies on the role of oxygen, and it is an often underemphasized aid, which the results suggest is fairly important. Surprisingly, summit centrality for sex (indicating male) was relatively low compared to no-summit centrality indicating that being male had low importance in the chances of success at summit. Lastly, the largest differences in summit vs no summit were from identity (sherpa were much more likely to succeed), and age ($<40$ year olds were much more likely to succeed), both of which are intuitive and also seen in previous studies \cite{huey2020}. 

\subsection{Expedition-wide factors}

While the intra-expedition network provides insight into personal features that determine success, expedition-wide factors also play an important role, particularly in globally cooperative expeditions. The expedition-wide factors considered here are: (1) number of days to summit from base camp, (2) number of high points/camps, (3) expedition size (including hired personnel), (4) ratio of number of paying climbers to number of hired personnel. 

The expedition success rate is defined as the fraction of climbers that succeed at summiting. The importance of an expedition-wide factor can be inferred from the correlation between the values of a given factor across a range of expeditions and their corresponding success rates. 

Fig. \ref{fig:centrality}(b) shows the Pearson's correlation coefficient between the expedition-wide factors and the success rate. A higher correlation implies higher importance in determining success. Despite sherpas having a high chance of personal success as seen in section \ref{sec:personal}, the ratio of number of paying members to number of hired personnel on the team is only weakly correlated with success, i.e., has a relatively small effect on \textit{expeditional} success. Both number of camps above basecamp and days to summit/high point are negatively correlated with success, as one might expect, with the latter having a larger effect. Also surprisingly, the \textit{expedition size} is found to be relatively important in determining success (with a correlation coefficient of $>$ 0.5). All p-values are extremely low, indicating that the correlation is statistically significant except for the number of members to hired personnel.

\section{Discussion}

This work presents the first network-based simplicial analysis of mountaineering data, studying the effect of the structure and strength of relationships on the nature of cooperation and success, both from an individual perspective and across the expedition. Using the Himalayan dataset, it establishes that relationships between climbers play an integral role by showing that the chances of summit failure (due to fatigue, logistical failure etc.) drastically reduce when climbing with repeat partners, especially for more experienced climbers. 
Further, individuals with high influence, i.e., belonging to a simplex with large simplicial dimension (encoding a previous joint expedition with a large number of members) were more likely to be successful in summiting, irrespective of experience level. However, the effects of having subgroup relationships on the collective group behavior varied. Specifically, expeditions with large simplexes that tended to have lower weight (indicating weak relationships) had a more cooperative style, with the average simplicial dimension being a good predictor of the expedition success rate. In contrast, expeditions with smaller simplexes of typically high weight encoded strong relationships between a small group of people, and tended to be more polarized. In such expeditions, individuals that were a part of the highly weighted simplex had a high likelihood of succeeding, whereas those that weren't had a low likelihood of success. 
It also studies various other indicators of success, such as personal features that are more important in individualistic expeditions, and expedition-wide factors are more important in cooperative expeditions.
A bipartite climber-feature network is projected into feature face to study the relative important of personal features. The largest difference in centralities amongst successful and unsuccessful groups is found in the `age' node, indicating that it's the strongest driver of success. The expedition-wide factors that have high correlation with expeditional success are high expedition size and low total number of days. 

In conclusion, this work presents novel analyses and new results that demonstrate the importance of different types of inter-personal relationships at high peaks on the extent of cooperation vs individualism, and effect on success, both from intra-group individual as well as expedition-wide perspectives. It extends work from \cite{krishnagopal2021success} studying the effects of both personal and expeditional factors on success, and highlights their relative importance in expeditions of different styles. Lastly, it is the first work applying simplicial complexes and topological data analysis to mountaineering data, opening it up for further analysis from the network science community.
Code can be found at \textit{https://github.com/chimeraki/mountaineeringsimplicial}.


\begin{backmatter}

	\section*{Availability of data and materials}
The datasets used in this paper in the Himalayan dataset \cite{data04} which is openly available at $www.himalayandatabase.com$

	\section*{Competing interests}
	The authors declare that they have no competing  interests.
	
	\section*{Funding}
	No external funding was obtained for this research.

\bibliographystyle{bmc-mathphys}    
\bibliography{himalaya}

\end{backmatter}  

\end{document}